\documentclass[seceq]{ptptex}

\usepackage{graphicx}
\usepackage{wrapft}

\title{
Absolute Rate, Evolving Luminosity Function, and
Evolving Jet Opening Angle Distribution for Long
Gamma-Ray Bursts 
}

\author{
Tatsushi \textsc{Matsubayashi},$^{1,}$\footnote{E-mail: 
tatsushi@riken.jp}
Ryo \textsc{Yamazaki},$^{2,3,}$\footnote{E-mail: 
ryo@vega.ess.sci.osaka-u.ac.jp}
Daisuke \textsc{Yonetoku},$^{4,}$\footnote{E-mail: 
yonetoku@astro.s.kanazawa-u.ac.jp}
Toshio \textsc{Murakami}$^{4,}$\footnote{E-mail:
murakami@astro.s.kanazawa-u.ac.jp} 
and 
Toshikazu \textsc{Ebisuzaki}$^{1,}$\footnote{E-mail: 
ebisu@riken.jp}
}

\inst{
$^1$Ebisuzaki Computational Astrophysics laboratory,  
    Institute of Physical and Chemical Research (RIKEN),
    Wako,  351-0198 Japan  \\
$^2$Department of Earth and Space Science, Osaka University, 
    Toyonaka 560-0043, Japan  \\
$^3$Department of Physics, Hiroshima University, Higashi-Hiroshima,
    739-8526, Japan  \\

$^4$Department of Physics, Kanazawa University,
    Kanazawa, 920-1192, Japan
}

\recdate{April 19, 2005}

\abst{
Using the luminosity function and the apparent GRB rate inferred from
the spectral peak energy-luminosity relation, 
we investigate the absolute GRB formation rate, taking into account
the effects of both the jet-luminosity evolution and 
the jet opening angle evolution.
In the case that there is no jet opening angle evolution,
the jet-corrected luminosity for high-redshift GRBs is much
larger than  typical values for low-$z$ ($z\sim1$) bursts.
On the other hand, if the jet-luminosity does not evolve with time,
the jet opening angle for high-$z$ bursts should be much smaller 
than that for low-$z$ GRBs.
Therefore, it is preferable to take into account both evolution 
effects.
We also estimate the local GRB event rate in a galaxy of
 $\sim 10^{-7}$--$10^{-5}$~yr$^{-1}$.
}

\begin{document}

\maketitle

\def\VVM{\langle V/V_{\rm max}\rangle}

\section{Introduction}
\label{sec1}

The luminosity function and the  formation rate history
of gamma-ray bursts (GRBs) are important for understanding
the nature of these events.
The formation history of GRBs provides constraints
on the population of progenitors.
If we know the absolute value of the GRB rate, we can study
their relation to other objects potentially related to GRBs.
It is fairly certain that  long GRBs arise 
from relativistic jets with typical opening half-angles of 
$\sim0.1$~rad\cite{meszaros02,zhang03}.
For this reason, determining
 the corrections due to jet collimation and/or
jet structure is essential in the study of  the GRB rate.
The luminosity function may directly reflect the properties of 
the progenitor, such as the jet structure, the jet kinematics, and so on.

GRBs could  be  useful tools to investigate  high redshift objects.
The connection between  long duration GRBs and supernovae is 
strongly suggested, and it is known that at least some GRBs arise 
from the collapse of  massive stars \cite{dv03,ga98,stanek03,hj03}.
The universe began to reionize at $z \sim 17$ \cite{spergel03}, 
which implies that the first generation 
objects had already been formed by at least $z \sim 20$.
Because they are massive, they immediately exploded and
 produced GRBs  at a redshift of $z \sim 20$.
Indeed, some of their prompt emissions and afterglows are bright enough 
to be actually detected\cite{lr00,cl00}.
The absolute rate of prompt GRB emissions may provide some information
concerning the cosmic star formation history \cite{totani97,murakami05}. 
Also, the afterglows of high-$z$ GRBs may provide information
concerning the cosmic reionization history 
\cite{iyn04,inoue04,ioka03,fl02,bl04}.

Many authors have attempted to derive the luminosity function and
absolute rate of GRBs.
Schmidt\cite{schmidt1999} derived 
the luminosity function and
the absolute value of the local GRB rate
so that $\VVM$ and the observed event rate of BATSE bursts are reproduced.
Later, he obtained the $\VVM$-hardness relation for BATSE GRBs that was
used  to determine the luminosity function and 
the absolute local GRB rate\cite{schmidt2001}.
The value he derived is $\sim0.5$~Gpc$^{-3}$yr$^{-1}$.
However, in that derivation, the geometrical corrections were not considered.
The luminosity function and the formation  rate history of long GRBs 
have been simultaneously derived using some luminosity indicators, such as
the luminosity-variability relation\cite{fenimore,lloyd02} and
the luminosity-lag relation\cite{norris,murakami03}.
Recently, it was found that the spectral peak energy is correlated with 
the peak luminosity with actually known distances of
up to $z\sim4.5$\cite{yonetoku2004}.
The correlation is tightest  in such kinds of luminosity 
indicators\cite{amati,atteia,ghirlanda2004}.
Using the peak energy-luminosity relation as a luminosity indicator, 
Yonetoku et al.\cite{yonetoku2004}
investigated the apparent GRB luminosity function (see \S~2.1)
and the GRB formation rate for redshifts $z\lesssim12$
and suggested that the
GRB formation rate rapidly increases up to $z \sim 1$ and then continues
to increase toward  higher redshifts $z\sim 12$. 
This behavior of the formation history is somewhat similar to those 
derived using other luminosity indicators\cite{fenimore,lloyd02,murakami03}.

There is evidence suggesting that the (isotropic-equivalent) luminosity 
evolution of GRBs takes the form $\propto(1+z)^{2.6}$\cite{yonetoku2004},
which is quite similar to that of quasars
\cite{caditz1990, maloney1999}. 
The luminosity evolution for GRBs has been independently reported 
\cite{lloyd02, wei03, firmani04, yonetoku2004}.
The origin of the luminosity evolution is not yet clear.
The jet-corrected luminosity,
$L_j=Lf_b$, where $f_b=\theta_j{}^2/2$ is the beaming factor
and $\theta_j$ is the jet opening half-angle, 
clusters very near the standard value\cite{lamb2005}.
The same is true for the jet-corrected 
energy\cite{frail2001,bloom2003,ghirlanda2004}.
These facts suggest that the luminosity evolution may result from 
either  jet-opening angle evolution\cite{lloyd02} or 
jet-luminosity evolution.

The effect of  jet collimation is investigated explicitly in
Ref.~\citen{guetta2005a}, where the luminosity function is studied
assuming that its functional form is a double power-law
and that the jet-corrected luminosity $L_j$ is constant.
For the structured jet case, general formulae to calculate 
the luminosity function and/or the local GRB rate
from the angular distribution of kinetic energies in the jet
has been presented\cite{perna2003,guetta2005b}.
One  crucial assumption in these works is that 
the apparent GRB rate is proportional to the cosmic star formation rate.
Furthermore, in Ref.~\citen{guetta2005a},
luminosity evolution is not considered.

In this paper, using the luminosity function and the apparent GRB rate 
derived in Ref.~\citen{yonetoku2004}, we study
the absolute GRB formation rate, taking into account the effects of
both the jet-luminosity evolution and the jet opening angle evolution.
A general formula to calculate the true GRB rate is presented.
The most important point in the derivation
is the geometrical correction of the jet opening angle, $\theta_{j}$.
Instead of using the distribution function for $\theta_j$, 
we consider an approach similar to that used in Ref.~\citen{guetta2005a}.
The fact that the jet-corrected luminosity
$L_j$ clusters in narrow ranges\cite{lamb2005}
enables us to calculate the jet-correction factors
without directly taking into account the opening angle distribution.
We also study the opening angle evolution.
If we assume that the origin of the luminosity evolution is the jet 
opening angle evolution, the high-redshift GRBs are collimated much 
more tightly than the nearby events. 
Throughout the paper, we assume a flat, isotropic universe with 
$\Omega_{M}=0.32, \Omega_{\Lambda}=0.68$, and $h=0.72$.

\section{The absolute GRB formation rate}
\subsection{Apparent luminosity function of GRBs}
The apparent luminosity function of GRBs, that is the rate of
actually detected GRBs per unit comoving volume at a redshift $z$ and a 
peak luminosity $L$, is written 
\begin{eqnarray}
\Phi (L,z) = 
\rho(z) \phi(L/g(z)) /g(z)~~,
\label{eq:Phi}
\end{eqnarray}
where $\rho(z)$, $g(z)$, and $\phi(L')$ are the GRB formation rate,
the luminosity evolution, and the local luminosity function,
respectively \cite{caditz1990, maloney1999,lloyd02}.
In this paper, we adopt the functional forms of these quantities
given below, as derived in Ref.~\citen{yonetoku2004}.
 
The GRB formation rate is given by
\begin{eqnarray}
\rho (z) = \rho_{0}
\left \{ 
\begin{array}{@{\,}ll}
 \left[(1+z)/2 \right]^{6.0} & 0<z<1~~,   \\
 \left[(1+z)/2 \right]^{0.4} & 1<z<z_{\rm max}~~, \\
 0        & z_{\rm max}<z~~,   
\end{array} \right.
\label{eq:rho}
\end{eqnarray}
where $\rho_{0}=\rho(z=1)$ 
is the normalization constant, and $z_{{\rm max}}$ is 
the epoch at which the first GRBs formed. 
Here we assume a broken power-law form. However,
the results obtained in the present paper are almost unchanged 
if  other smooth functions are adopted.

The luminosity evolution is assumed to take the form
\begin{eqnarray}
g(z) = {\left( \frac{1+z}{2} \right)}^{\alpha}~~,
\label{eq:g(z)}
\end{eqnarray}
where, throughout the paper, we adopt $\alpha =2.6$,
as derived in Ref.~\citen{yonetoku2004}.

The local luminosity function $\phi(L')$,
where $L'\equiv L/g(z)$ represents the peak luminosity after
removing the luminosity evolution effect,
 is derived from the universal
cumulative luminosity function $\psi (L')$ as
\begin{eqnarray}
 \phi(L') = - \frac{d}{dL'}\psi (L') ~~,
 \label{eq:phi}
\end{eqnarray}
and we assume the following form of $\psi (L')$
in the range $L'_{\rm min} <L' <L'_{\rm max}$:
\begin{eqnarray}
\psi (L') = \psi_{0}
 \left({L' / L'_{\ast}}\right)^{A}
 {\left[ 1+ \left({L' / L'_{\ast}}\right)^{s}\right]}^{(B-A)/s} ~~.
\label{eq:psi}
\end{eqnarray}
Here $A$ and $B$ are the power law indices in the asymptotic forms,
and $s$ describes the smoothness of the transitions between the high and
low $L'$ regimes.
We set these parameters as $A=-0.29$, $B=-1.02$, and $s=2.6$, 
in order to reproduce the results presented in Fig.~6 of Ref.~\citen{yonetoku2004}. 
Therefore, the break point of $\psi(L')$ is
$L'_{\ast}=1.1\times2^{\alpha}\times10^{51}$~erg~s$^{-1}$, because the functional
form of $g(z)$ adopted in this paper has an extra factor of $2^\alpha$
compared with that used in Ref.~\citen{yonetoku2004}. 
Finally, the normalization constant $\psi_0$ is chosen so as to satisfy
\begin{equation}
\int_{L'_{\rm min}}^{L'_{\rm max}}\phi(L')dL'=1~~.
\label{eq:normalization}
\end{equation}

As suggested in Ref.~\citen{yonetoku2004},
we assume the universality of the local luminosity function $\phi(L')$. 
Then, both $L'_{\rm max}$ and $L'_{\rm min}$ should be independent of
$z$; that is, we can write 
\begin{eqnarray}
\left\{ \begin{array}{@{\,}ll}
L_{{\rm max}}(z) = L'_{{\rm max}} \times g(z)~~,\\
L_{{\rm min}}(z) = L'_{{\rm min}} \times g(z) ~~.
\end{array} \right. 
\label{eq:LmaxLmin}
\end{eqnarray}
Hence both the maximum and the minimum values of the luminosity 
evolve with $g(z)$. In the fiducial case, the values 
$L'_{\rm max}=1\times10^{54}$~erg~s$^{-1}$ and 
$L'_{\rm min}=1\times10^{49}$~erg~s$^{-1}$ 
are adopted\cite{yonetoku2004}.

We determine the normalization constant $\rho_0$ as follows.
The actual event rate of BATSE-observed bursts, with observed peak
fluxes larger than $F_{\rm obs}$, is given by
\begin{eqnarray}
N_{\rm obs}(>F_{\rm obs}) 
= \zeta_{\rm exp}
\int_{0}^{z_{\rm lim}} dz \,
\frac{1}{1+z} \frac{dV}{dz} 
\int_{L_{\rm lim}(z)}^{L_{\rm max}(z)} dL\,
\Phi (L,z) ~~,
\label{eq:normalize}
\end{eqnarray}
where $V$, $L_{\rm lim}$, and
$z_{\rm lim}$ are the comoving volume, the limiting flux for given $z$, 
and the upper bound on the redshift, respectively.
The mean exposure efficiency $\zeta_{\rm exp}$ is approximately 48\% 
for the BATSE experiment \cite{paciesas1999}. 
The limiting flux can be written as 
\begin{eqnarray}
L_{\rm lim}(z) =4\pi d_{L}(z)^{2} \times F_{\rm obs}, 
\label{eq:limitflux}
\end{eqnarray}
where $d_{L}(z)$ is the luminosity distance. 
In Ref.~\citen{yonetoku2004},  689 BATSE bursts were considered.
Their observed peak fluxes in the 30--10000~keV band are
larger than $2\times10^{-7}$~erg~cm$^{-2}$s$^{-1}$ and
their redshifts are smaller than $12 ( = z_{\rm lim})$.
The corresponding observation period is 9.0~yrs, because they selected the GRB
sources from ID\# 105(1991.111) to ID\# 8121(2000.147) 
(see Table 2 of Ref.~\citen{yonetoku2004}).
Hence we find $N_{\rm obs}(>2\times10^{-7}$~erg~cm$^{-2}$s$^{-1})$
$\sim 76.6$~events~yr$^{-1}$.
We can determine the value of the normalization constant $\rho_{0}$ 
needed to reproduce this value.
The results are shown in Table~\ref{table:LmaxLmin}.
It is found that $\rho_0$ changes by a few percent if $L'_{\rm max}$ 
ranges between $1\times 10^{53}$ and $1\times 10^{55}$~erg~s$^{-1}$.
The dependence of $L'_{\rm min}$ is estimated as follows.
Equation~(\ref{eq:normalization}), together with
Eqs.~(\ref{eq:phi}) and (\ref{eq:psi}), gives the approximate form
$\psi_0\sim(L'_{\rm min}/L'_*)^{-A}$. Then,
Eq.~(\ref{eq:normalize}) reads 
$\rho_0\propto\psi_0{}^{-1}\propto(L'_{\rm min}/L'_*)^{A}$.
Therefore, as shown in Table~\ref{table:LmaxLmin},
when $L'_{\rm min}$ becomes 10 times larger (smaller),
 $\rho_0$ becomes about 2~times smaller (larger).
Finally, we find that $\rho_0$ changes by a factor of 2
when the parameters in the luminosity function
are varied in ranges corresponding to 90\% 
of the uncertainties on the observational results
derived in Ref.~\citen{yonetoku2004}.

\begin{table}[t]
\begin{center}
\caption{
The values of $\rho_0=\rho(z=1)$ (in units of ${\rm Gpc^{-3}~yr^{-1}}$)
for various luminosity ranges [see Eq.~(\ref{eq:rho})].}
\label{table:LmaxLmin}
\begin{tabular}{cc|ccc}
\hline
\hline
&&& $L'_{{\rm max}}{}^{\rm a)} $ & \\
&& $1\times10^{53}$& $1\times10^{54}$& $1\times10^{55}$ \\
\hline
                            & $1\times10^{48}$ & 1.69 & 1.63 & 1.62 \\
$L'_{{\rm min}}{}^{\rm a)} $ & $1\times10^{49}$ & 0.86 & 0.84 & 0.83 \\
                            & $1\times10^{50}$ & 0.44 & 0.43 & 0.43 \\
\hline
\end{tabular}
\end{center}
\begin{center}
{\footnotesize
a) The maximum and minimum luminosities in units of 
           ${\rm erg~s^{-1}}$ [see Eq.~(\ref{eq:LmaxLmin})].  \\
}
\end{center}
\end{table}

We obtain $\rho(0)\simeq0.01$~Gpc$^{-3}$yr$^{-1}$
and $\rho(1)\simeq1.0$~Gpc$^{-3}$yr$^{-1}$ in this paper.
Let us compare these with previous results. From
the results obtained in Ref.~\citen{firmani04}, in which
 the functional form of $\rho(z)$ is determined
so as to reproduce both the observed peak flux and redshift 
distributions, 
the values $\rho(0)\simeq0.04$--0.4~Gpc$^{-3}$yr$^{-1}$
and $\rho(1)\simeq0.2$--3.8~Gpc$^{-3}$yr$^{-1}$ can be inferred.
We thus find that our results are roughly consistent with 
those derived in Ref.~\citen{firmani04}. 
However, Schmidt\cite{schmidt2001} estimated the local GRB rate density
to be $\rho(0)\sim0.5~{\rm Gpc^{-3}yr^{-1}}$, which is more than
one order of magnitude higher than our value.
This difference may result from an insufficient amount of data 
and low resolution in the redshift range $0<z<1$. 
The form of $\rho(z)$ we employ might be steeper than the actual one.
Therefore, because the function $\rho(z)$ is 
normalized at $z=1$ in this paper,  
the local GRB rate that we derived at $z=0$ might be smaller than
the actual value. 

\subsection{GRB formation rate corrected by the jet opening angle}

We consider the {\it true} rate of GRBs which actually occurred
in the redshift range $z_1<z<z_2$.
For this purpose, it is necessary to consider
the geometrical correction of the jet opening angle, $\theta_{j}$.
Although the jet structure is still unknown, 
we employ a uniform jet model.
Taking into account the correction due to jet collimation,
the cumulative GRB rate is calculated as
\begin{eqnarray}
N(z_1,z_2) 
&=& 
\int_{z_1}^{z_2} dz \, \frac{1}{1+z} \frac{dV}{dz}
\int_{L_{\rm min}(z)}^{L_{\rm max}(z)} dL\, \Phi (L,z) 
\nonumber \\
&& \ \ \times
\int dL_j \, P(L_j,z)\,f_b(L,L_j)^{-1}~~,
\label{eq:rate_corrected}
\end{eqnarray}
where $L_j$ and $f_b$ are the peak luminosity confined in the jet and
the beaming factor, given by
\begin{eqnarray}
f_b(L,L_j)&=&1-\cos\theta_j(L,L_j) \nonumber \\
      &=&\frac{L_j}{L} \ (<1)~~,
\label{eq:beam}
\end{eqnarray}
respectively.
The quantity $L_j$ is assumed to be  clustered in a narrow range 
around the standard value\cite{lamb2005},
and its distribution is approximately log-normal:
\begin{eqnarray}
P(L_{j},z) =
\frac{1}{\sqrt{2\pi}\sigma L_{j}}
\exp
\left[ -\frac{1}{2\sigma^{2}}
\left(\log L_{j} - \log \tilde{L_j}(z)\right)^{2}
\right] ~~,
\label{eq:lognormal}
\end{eqnarray}
where $\log \tilde{L_{j}}(z)$ and $\sigma$ are the mean and  standard 
deviation of the luminosity distribution, respectively.
In this paper, the evolution of the jet-corrected luminosity is
assumed to take the simple form
\begin{eqnarray}
\tilde{L_j}(z)
= L_{j1} \left(\frac{1+z}{2}\right)^{\beta}~~,
\label{eq:Lj_mean}
\end{eqnarray}
where $\beta$ is the slope index of the jet-luminosity evolution. 
We  discuss in \S\ref{sec:discussion}
the validity of our assumed functional forms
of $P(L_j,z)$ and $\tilde{L_j}$.

Substituting Eqs.~(\ref{eq:Phi}), (\ref{eq:beam}), (\ref{eq:lognormal}), 
and (\ref{eq:Lj_mean}) into Eq.~(\ref{eq:rate_corrected}), 
we obtain the cumulative GRB rate as 
\begin{eqnarray}
N(z_1,z_2) &=& 
\int_{z_1}^{z_2} dz \, \frac{1}{1+z} \frac{dV}{dz}
\rho(z)\left< f_{b}^{-1}\right>_z~, 
\label{eq:rate_corrected2}
\end{eqnarray}
where 
\begin{eqnarray}
\left< f_{b}^{-1}\right>_z 
&\equiv&
\frac{\int_{L_{\rm min}(z)}^{L_{\rm max}(z)}dL\,\Phi(L,z) \, 
L /\left< L_{j}\right>_z}
{\int_{L_{\rm min}(z)}^{L_{\rm max}(z)}dL\,\Phi(L,z)} 
\nonumber \\
&=& 
\frac{10^{\frac{1}{2}\sigma^2\ln10}}{L_{j1}}
\left( \frac{1+z}{2}\right)^{\alpha - \beta}
\int_{L'_{\rm min}}^{L'_{\rm max}}dL'\,\phi(L') \, L'
\label{eq:beam_mean}
\end{eqnarray}
is the luminosity weighted beaming factor, and
\begin{eqnarray}
{\left< L_j\right>}_{z} & \equiv &
\left[
\int dL_{j} \,P(L_{j},z) L_j^{-1}
\right]^{-1} \nonumber \\
&=&
\frac{L_{j1}}{10^{\frac{1}{2}\sigma^2\ln10}}
\left( \frac{1+z}{2}\right)^{\beta}
\label{eq:meanlumi}
\end{eqnarray}
is the mean jet-luminosity.

We choose the value of 
$\left< L_j\right>_{z=1}=L_{j1}/10^{\frac{1}{2}\sigma^2\ln10}$
so that $\left< f_b^{-1}\right>_z$ is 200 at $z=1$.
In this case, the mean jet opening half-angle, defined by
$\left< \theta_{j}\right>_{z} \equiv[\left< f_b^{-1}\right>/2]^{-1/2}$, 
is equal to the typical value of 0.1~rad at $z=1$.
Then we find 
\begin{eqnarray}
\left< L_j\right>_{z=1}&\equiv&
\frac{L_{j1}}{10^{\frac{1}{2}\sigma^2\ln10}} \nonumber\\
&=&
2.5\times 10^{49}~{\rm erg}~{\rm s}^{-1}
\left(\frac{\left< f_b^{-1}\right>_{z=1}}{200}
\right)^{-1}
\label{eq:Ljnormalize}
\end{eqnarray}
with the fiducial parameters
($L'_{\rm max}=1\times10^{54}$~erg~s$^{-1}$ 
 and $L'_{\rm min}=1\times10^{49}$~erg~s$^{-1}$).
The value of $\left< L_j\right>_{z=1}$ changes by only
a factor of 3 when $L'_{\rm max}$ and $L'_{\rm min}$ vary 
between $1\times 10^{53}$ and $1\times 10^{55}$~erg~s$^{-1}$ and 
between $1\times 10^{48}$ and $1\times 10^{50}$~erg~s$^{-1}$, respectively.
Substituting Eq.~(\ref{eq:Ljnormalize}) into 
Eq.~(\ref{eq:beam_mean}), we derive
\begin{equation}
\left< f_{b}^{-1}\right>_z = 200
\left( \frac{1+z}{2}\right)^{\alpha - \beta}
\left(\frac{\left< f_b^{-1}\right>_{z=1}}{200}\right)~~,
\end{equation}
and 
\begin{equation}
\left< \theta_{j}\right>_{z}=0.1
\left( \frac{1+z}{2}\right)^{\gamma}
\left(\frac{\left< f_b^{-1}\right>_{z=1}}{200}\right)^{-1/2}~~,
\label{eq:angleevolution}
\end{equation}
where $\gamma=(\beta-\alpha)/2$.

\begin{figure}[t]
\parbox{\halftext}{
\begin{center}
\includegraphics[scale=.7]{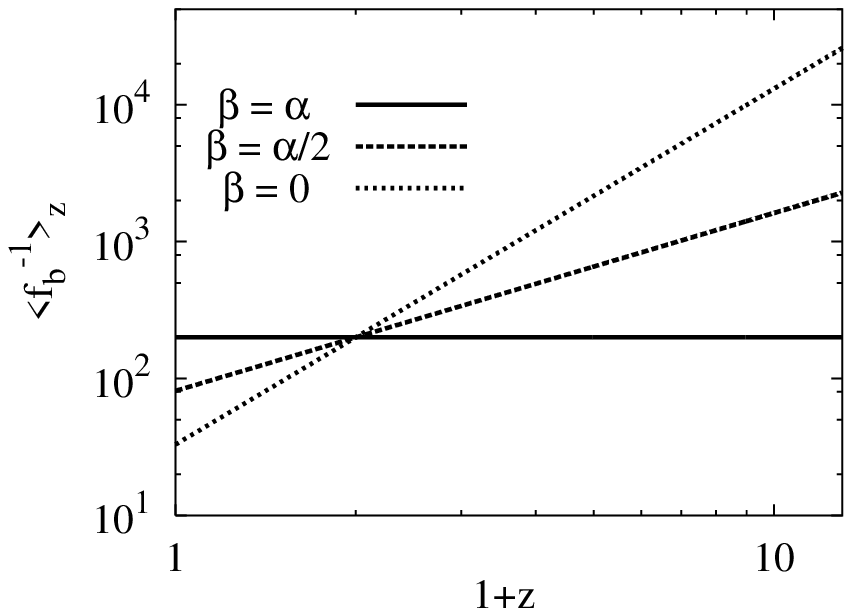}
\end{center}
\caption{ 
Luminosity-weighted beaming factor $\left< f_{b}^{-1}\right>$
as a function of the source redshift $z$.
The solid, dashed, and dotted lines correspond to 
$\beta = \alpha$, $\alpha/2$, and $0$, respectively. 
In the case $\beta=\alpha$ (solid line), 
the beaming factor is independent of $z$,
with the value $\left< f_b^{-1}\right> = 200$.
}
\label{fig:beaming}
}
\hfill
\parbox{\halftext}{
\begin{center}
\includegraphics[scale=.7]{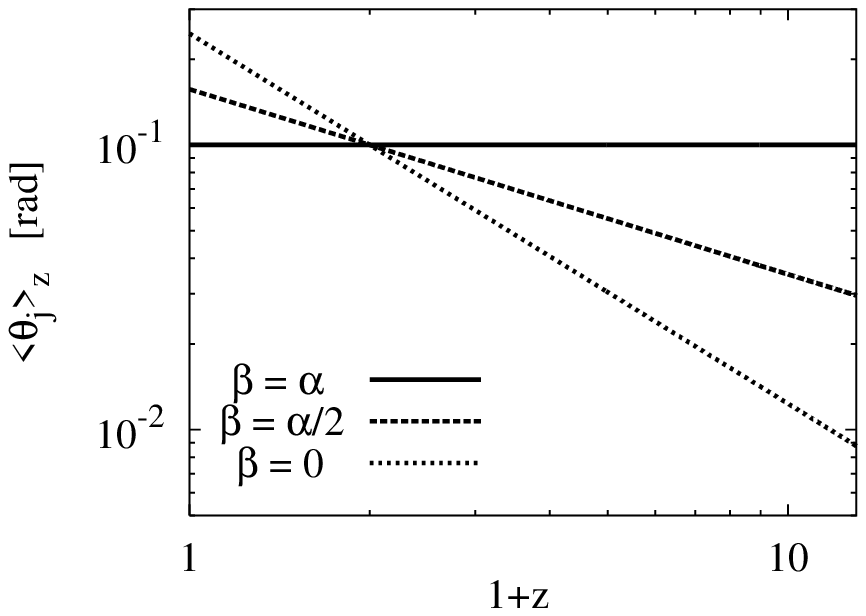}
\end{center}
\caption{ 
Mean opening half-angle $\left< \theta_j\right>$ 
as a function of the source redshift $z$.
The solid, dashed, and dotted curves correspond to 
$\beta = \alpha$, $\alpha/2$, and $0$, respectively. 
In the case $\beta=\alpha$ (solid line), 
the mean jet opening angle is independent of $z$,
with the value $\left< \theta_{j}\right> = 0.1$~rad.
} 
\label{fig:theta}
}
\end{figure}

\begin{wraptable}{r}{\halftext}
\caption{Models.}
\label{table:beta}
\begin{center}
\begin{tabular}{c|c}
\hline
\hline
$\beta$ & Evolution \\
\hline
\shortstack{~\vspace{5pt}\\
$\alpha $ \\
~\vspace{5pt}}& 
\shortstack{
{\bf Jet-luminosity evolution }\\ 
({\it \small no jet-opening angle evolution}) } \\
\shortstack{~\vspace{5pt}\\ $\alpha /2 $ \\~\vspace{5pt}} & 
\shortstack{ ~\vspace{5pt}\\{\bf Inter-mediate }\\~\vspace{5pt} }\\
\shortstack{~\vspace{5pt}\\ 0 \\ ~\vspace{5pt} }& 
\shortstack{
{\bf Jet-opening angle evolution}\\ 
({\it \small no jet-luminosity evolution})} \\
\hline
\end{tabular}
\end{center}
\end{wraptable}
Figures~\ref{fig:beaming} and \ref{fig:theta} display the 
luminosity-weighted beaming factor, $\left< f_b^{-1}\right>_{z}$, 
and the mean opening half-angle, $\left<\theta_j\right>_{z}$, 
as functions of $z$.
The value of $\beta$ is uncertain.
As seen in Table~\ref{table:beta},
 we consider three cases: 
$\beta=\alpha$ ({\it no jet-opening angle evolution}),
$\beta=\alpha/2$ ({\it intermediate}), 
and $\beta=0$ ({\it no jet-luminosity evolution}).
In the case $\beta=0$, we find
$\left< f_b^{-1}\right>_{z}\sim 3\times10^4$ 
and $\left< \theta_j\right>_{z}\sim 9\times 10^{-3}$ 
for $z\sim12$ in the fiducial case.
The latter is about $10$ times smaller than the typical
value for $z\sim1$.
By contrast, when $\beta=\alpha=2.6$, 
the beaming factor is independent of $z$, and we have
 $\left< f_b^{-1}\right> = 200$.
However, the mean jet-corrected luminosity $\tilde{L}_j$
increases with $z$ [see Eq.~(\ref{eq:Lj_mean})]
and $\tilde{L}_j(12)\sim 1.3 \times10^2\tilde{L}_j(1)$,
which is also unnatural.
Hence,   intermediate values of $\beta$, in the neighborhood of
$\beta\sim\alpha/2$,
for which both $\tilde{L}_j$ and $\left< f_b^{-1}\right>$ depend on $z$,
may be preferable.

\begin{figure}[t]
\parbox{\halftext}{
\begin{center}
\includegraphics[scale=.7]{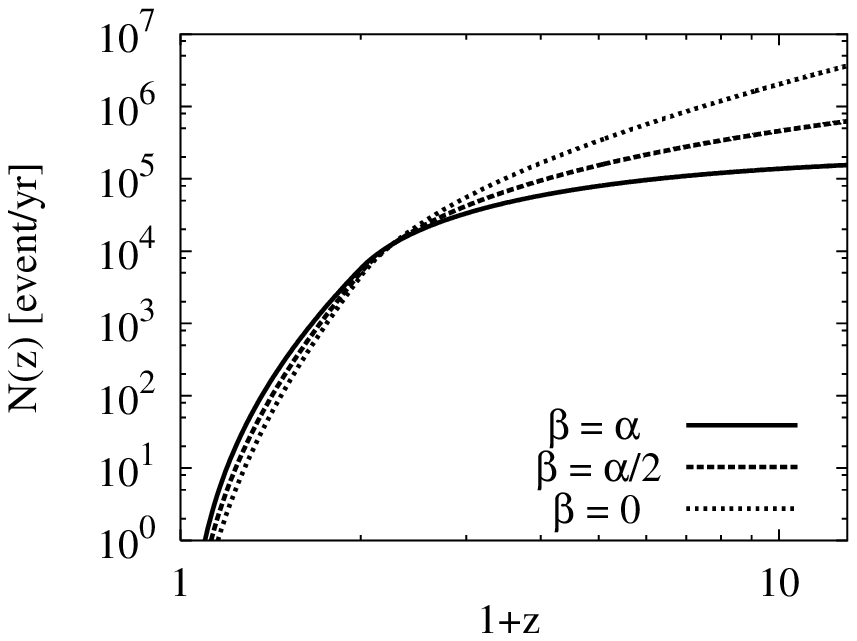}
\end{center}
\caption{ 
The cumulative GRB rate $N(0,z)$ [see Eq.~(\ref{eq:rate_corrected2})]. 
The solid, dashed, and dotted curves correspond to
$\beta = \alpha$, $\alpha/2$, and $0$, respectively. 
These are frequencies of GRBs that actually occurred at redshifts from $0$ to $z$. 
}
\label{fig:cum_true}
}
\hfill
\parbox{\halftext}{
\begin{center}
\includegraphics[scale=.7]{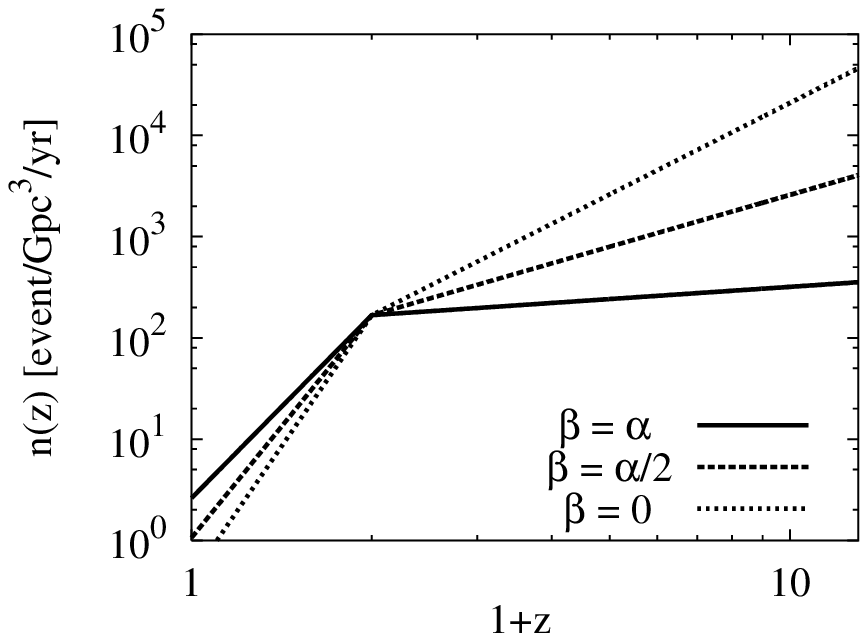}
\end{center}
\caption{ 
The comoving GRB rate density $n(z)$.
The solid, dashed, and dotted curves correspond to
$\beta = \alpha$, $\alpha/2$, and $0$, respectively. 
In the case $\beta=\alpha$, $n(z)$ is simply proportional to $\rho(z)$.
} 
\label{fig:comoving_rate}
}
\end{figure}

Figure~\ref{fig:cum_true} plots the cumulative GRB rate $N(0,z)$ 
as a function of $z$, which is the GRB frequency  at redshifts 
smaller than $z$. 
When we vary $\beta$ from $\alpha$ to $0$, $N(0, z_{{\rm max}})$ 
increases by approximately one order of magnitude,
and the deviation becomes larger for higher redshift.

Finally, the {\it true} comoving GRB rate density is given by
\begin{eqnarray}
n(z) &=&
(1+z) \left(\frac{dV}{dz}\right)^{-1}\frac{d}{dz}N(0,z) \nonumber \\
&=& \rho(z) \left< f_b{}^{-1}\right>_z~~.
\label{eq:n}
\end{eqnarray}
Figure~\ref{fig:comoving_rate} presents the results. 
In the case $\beta=\alpha$, the value of $\left< f_b^{-1}\right>_{z}$ 
does not depend on $z$, and thus $n(z)$ is simply proportional to $\rho(z)$.
It is also seen that 
when the jet opening angle evolution is considered ($\beta < \alpha$),
the GRB rate at high redshift becomes much larger than
that at low-$z$.

\section{Discussion}
\label{sec:discussion}

We have estimated the absolute GRB rate taking into account the effects of
 jet collimation. The 
normalization constant, $\rho_0$, was chosen so as to reproduce
 the BATSE detection rate
of bright GRBs with observed fluxes larger than 
$2\times10^{-7}$~erg~cm$^{-2}$s$^{-1}$.
In this paper, we specifically considered for the first time both the
jet luminosity evolution and the jet opening angle evolution.
Both evolution effects are important because when only one of them is 
taken into account, extreme results are obtained in the high redshift
regime, say, $z>10$.
When the mean jet opening angle does not depend on $z$ 
($\beta=\alpha$; {\it no jet-opening angle evolution}),
the jet-corrected luminosity $\tilde{L}_j$ for high-redshift GRBs is much
larger than the typical values for $z\sim1$ bursts.
On the other hand,
when $\tilde{L}_j$ is independent of $z$ 
($\beta=0$; {\it no jet-luminosity evolution}), 
the jet opening angle 
for high-$z$ bursts should be much smaller than that for low-$z$ GRBs.
Therefore, the intermediate case ($\beta\sim\alpha/2$), 
in which both $\tilde{L}_j$ and the jet opening angle varies with $z$,
is most preferable.
It is also found that for any value of $\beta$,
the true GRB rate $n(z)$ does not decrease for high-$z$
as long as $z<12$, and might not be simply proportional to
the cosmic star formation rate.
These facts might imply that the metalicity affects the properties of
the GRB progenitor.
For example, in the high-redshift epoch, 
because of the low metalicity,
the opacity of a star was small and the stellar envelope 
may not be blown off completely. 
Then, it might be difficult for the GRB jets to break out of 
the stellar surface, resulting in small opening angles.

\begin{figure}[t]
\parbox{\halftext}{
\begin{center}
\includegraphics[scale=0.25,angle=-90]{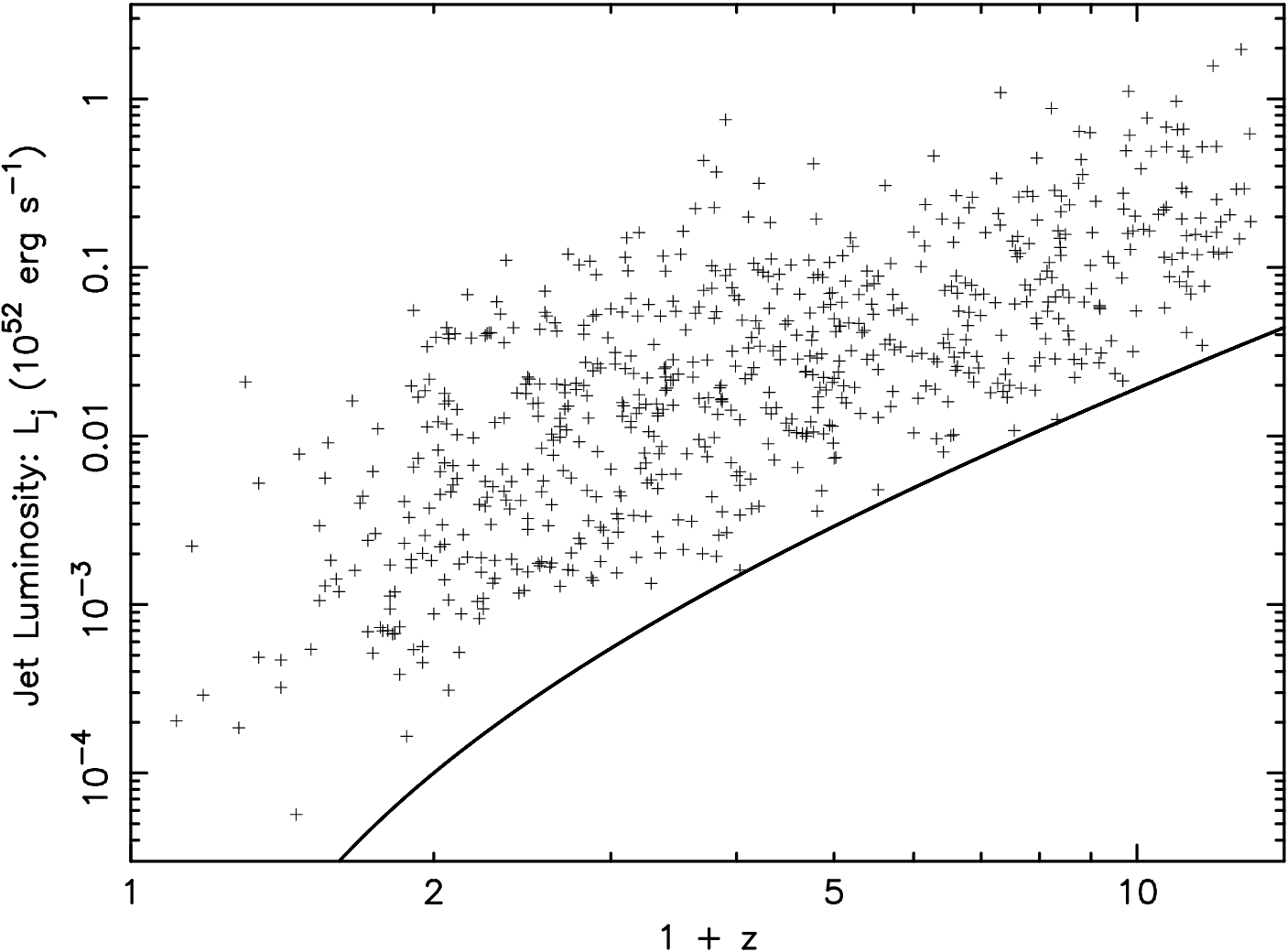}
\end{center}
\caption{ 
Distribution of the jet-corrected luminosity 
$L_j=f_bL$ (with $f_b=1-\cos\theta_j$, where $\theta_j$ is the jet
opening half angle) as a function of the redshift. The
redshift is estimated using the spectral peak energy-peak luminosity
relation (the Yonetoku relation)\cite{yonetoku2004}, 
and $\theta_j$ is estimated\cite{yonetoku2005} using 
the Yonetoku relation and the spectral peak 
energy-jet corrected energy relation 
(the Ghirlanda relation)\cite{ghirlanda2004}.
The solid curve represents the truncation of the upper bound of $L_j$
that is caused by the flux limit 
$F_{\rm lim}=1\times10^{-6}$~erg~cm$^{-2}$s$^{-1}$.
}
\label{fig:Lj_z}
}
\hfill
\parbox{\halftext}{
\begin{center}
\includegraphics[scale=0.25,angle=-90]{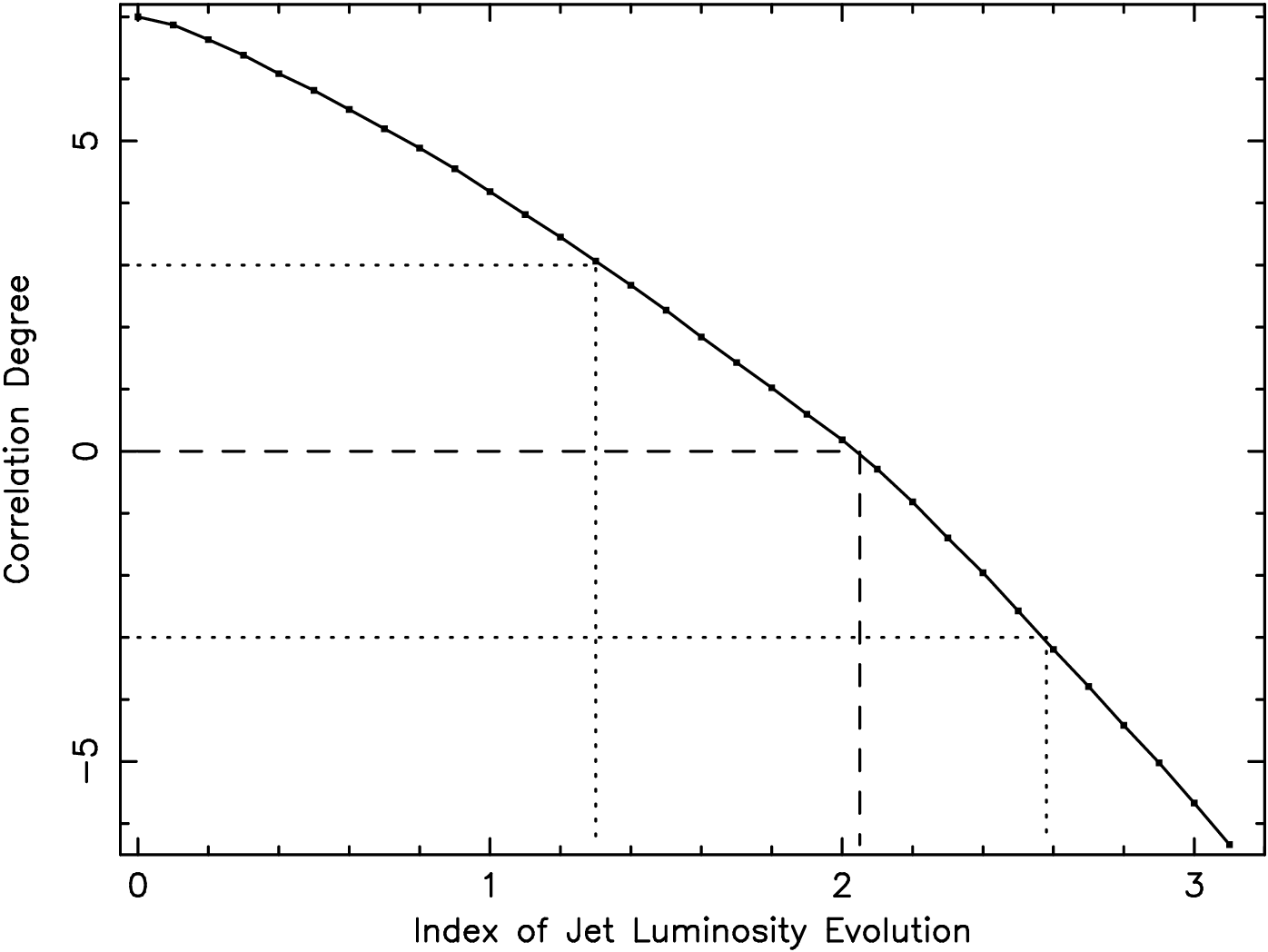}
\end{center}
\caption{ 
The best index of the jet luminosity evolution measured using the
$\tau$-statistical method.
When we assume $L_j$ evolution as a function of $(1+z)^\beta$, 
$\beta=2.05$ is the best value, at which the data correction degree
becomes zero. We also plot upper and lower bounds of the index
corresponding to 3$\sigma$ from the best value.
These bounds are given by $\beta=2.58$ and $\beta=1.30$.
The value $\beta=0$ lies approximately 8$\sigma$ from the best value,
and hence we reject this  case in which $L_j$-evolution
does not exist.
} 
\label{fig:index}
}
\end{figure}

Recently, the existence of opening angle evolution
of the form $\propto(1+z)^{\gamma}$ with $\gamma=-0.45^{+0.20}_{-0.18}$ 
was suggested in Ref.~\citen{yonetoku2005}.
Combining this with the result $\alpha=2.6^{+0.15}_{-0.20}$
derived in Ref.~\citen{yonetoku2004}, we find 
$\beta=\alpha+2\gamma=1.70^{+0.43}_{-0.41}$,  and thus
$\beta\simeq\alpha/2$.
The jet luminosity evolution is seen more clearly in 
Fig.~\ref{fig:Lj_z}.
Here we use the  data set analyzed in Ref.~\citen{yonetoku2005}.
Also, the redshifts and opening angles are estimated in the same manner
as in Ref.~\citen{yonetoku2005}. In this way, we can obtain 
the jet-collected luminosity $L_j$ as a function of $(1+z)$.
Using the generalized tau statistical method
\cite{petrosian1993,maloney1999b,lloyd02,yonetoku2004,yonetoku2005}, 
we obtain $\beta=2.05^{+0.53}_{-0.75}$ (see Fig.~\ref{fig:index}), 
which is consistent with the value $\beta=1.70^{+0.43}_{-0.41}$,
which is predicted using the combination of the luminosity evolution and
the jet opening angle evolution. 
Furthermore, Fig.~\ref{fig:Lj_z} shows that $L_j$ is
clustered in a narrow range around the mean value for each $z$ 
and that the distribution of $L_j$ for a given bin of $1+z$ 
roughly takes the form of a log-normal with a logarithmic mean 
$\sigma$ of approximately 0.6.
[When we adopt $\sigma\sim0.6$ and $\left< f_{b}^{-1}\right>_{z=1}=200$,
we find $L_{j1}\sim3\times10^{49}$~erg~s$^{-1}$ from 
Eq.~(\ref{eq:Ljnormalize}).]
We believe that $\sigma$ is independent of $z$, as assumed in this
paper, but, the lack of statistics prevents us from carrying out a
more detailed study of this point.
Of course, our method to estimate $L_j$ contains many uncertainties,
and hence more direct estimations of $z$ and $\theta_j$ are necessary.
If the {\it Swift} satellite collects data from many events with 
spectroscopically measured redshifts and jet opening angles determined 
from the wavelength-independent achromatic break in the afterglow light curves,
our understanding of evolution effects will be tested.

Using the local GRB rate, the event rate per galaxy can be estimated as
\begin{eqnarray}
r_{\rm GRB}&=&n(z)/n_{\rm gal} \nonumber \\
&\simeq&1\times10^{-5}\ {\rm yr}^{-1}\ {\rm galaxy}^{-1}
\left(\frac{n(z)}{10^2\ {\rm yr}^{-1}\ {\rm Gpc}^{-3}}\right)
\left(\frac{n_{\rm gal}}{10^{7}\ {\rm Gpc}^{-3}}
\right)^{-1}~~,
\end{eqnarray}
where $n_{\rm gal}$ is the total number density of spiral and 
elliptical galaxies,
which is typically $\sim1\times 10^{7}~{\rm Gpc^{-3}}$ \cite{mcmillan2001}.
In the case $\beta=\alpha/2$, we obtain
$r_{\rm GRB}=2\times10^{-5}$~yr$^{-1}$~galaxy$^{-1}$ for $z=1$
and $r_{\rm GRB}=1\times10^{-7}$~yr$^{-1}$~galaxy$^{-1}$
for $z=0$.
Sokolov\cite{sokolov2000} independently estimated the result
$r_{\rm GRB}\sim5\times10^{-8}\left< f_b{}^{-1}\right>$~yr$^{-1}$~galaxy$^{-1}$,
which is roughly consistent with our result.
Our values are comparable to the local rate of core-collapse 
(energetic Type~Ib/Ic) SNe,
 $\sim 10^{-6}$--$10^{-5}$yr$^{-1}$~galaxy$^{-1}$
\cite{podsiadlowski2004},
although the rate for type Ib/Ic SN at high redshift
is not known.
Even if collimated ultra-relativistic outflow and  relativistic
beaming effect prevent us from observing prompt GRB emissions far from
the jet axis, the emissions from associated energetic 
core-collapse SNe are nearly isotropic.
Hence, the expected SN rate can be estimated as
$N(0,z_{\rm SN})$.
The {\it Supernova/Acceleration Probe}\footnote{
{\ttfamily http://snap.lbl.gov/} }
will detect SNe with redshifts of up to $z_{\rm SN}\sim1.7$,
and thus we expect $N(0,z_{\rm SN}=1.7)\sim3\times10^4$~events~yr$^{-1}$.

\section*{Acknowledgments}
This work was supported in part by
Grants-in-Aid for Scientific Research 
of the Japanese Ministry of Education, Culture, Sports, Science
and Technology [No.~09245 (RY) and 15740149 (DY)].
TM is supported by the Junior Research Associate Program at
RIKEN (The Institute of Physical and Chemical Research), Japan. 

%


\end{document}